\documentclass[pre,twocolumn,showpacs,superscriptaddress,repreint]{revtex4-1}

\usepackage{graphicx}
\usepackage{dcolumn}
\usepackage{bm}
\usepackage{amsmath}
\usepackage{hyperref}
\usepackage{color}

\makeatother
\usepackage{tikz}
\begin{document}

\title{Transport of underdamped active particles in ratchet potentials}

\author{Bao-quan  Ai} \email[Email: ]{aibq@scnu.edu.cn}
\author{Feng-guo Li}
\affiliation{Guangdong Key Laboratory of Quantum Engineering and Quantum Materials, School of Physics and Telecommunication
Engineering, South China Normal University, Guangzhou 510006, China.}

\date{\today}

\begin{abstract}
\indent We study the rectified transport of underdamped active noninteracting particles in an asymmetric periodic potential. It is found that the ratchet effect of active noninteracting particles occurs in a single direction (along the easy direction of the substrate asymmetry) in the overdamped limit. However, when the inertia is considered, it is possible to observe reversals of the ratchet effect, where the motion is along the hard direction of the substrate asymmetry. By changing the friction coefficient or the self-propulsion force, the average velocity can change its direction several times.  Therefore, by suitably tailoring the parameters, underdamped active particles with different self-propulsion forces can move in different directions and can be separated.
\end{abstract}


\maketitle

\section{Introduction}
\indent  Recently, a class of active ratchet systems has been realized through the use of active matter, which are self-propelled
units that can be biological or non-biological in nature \cite{Reichhardt0}. Differently from passive Brownian particles, active particles, also known as self-propelled Brownian particles or microswimmers and nanoswimmers, are capable of taking up energy from their
environment and converting it into directed motion \cite{Bechinger}. Active ratchet effects and variations upon them will be a growing
field of research as the ability to fabricate additional types of artificial swimmers, nanobots, and other self-driven systems \cite{Bechinger}.

 \indent  Ratchets have been studied and experimentally realized for a variety
of systems including self-propelled particles on asymmetric substrates \cite{Galajda,Kaiser1,Koumakis0,Schwarz-Linek,Bricard,Mijalkov1,Kummel}.
 Rectification effects in active matter systems were first observed for run-and-tumble swimming bacteria moving through
an array of funnel-shaped barriers\cite{Galajda}. Subsequently, some theoretical and numerical studies focus on rectification of self-propelled particles \cite{Wan,Ghosh,Ghosh1,potosky,angelani,angelani2,angelani1,Potiguar,Koumakis,McDermott,Sandor,Lambert,Drocco,Li1,Li,Reichhardt2,Reichhardt1,Berdakin,Ai1,Ai,Mijalkov,Fily,Chen}.
A simple model of point-like run-and-tumble particles moving in a 2D container\cite{Wan} was studied with the same funnel barrier geometry used in the experiments \cite{Galajda}. Ghosh et. al. \cite{Ghosh} performed simulations of active Janus particles in an asymmetric channel and found that the rectification can be orders of magnitude stronger than that for ordinary thermal potential ratchets. Potosky and co-workers \cite{potosky}found that the spatially modulated self-propelled velocity can induce the directed transport. Angelani and coworkers \cite{angelani} studied self-driven particles in the presence of asymmetric piecewise periodic potentials. The collective ratchets and current reversals of active particles were observed in quasi-one-dimensional asymmetric substrates \cite{McDermott}.  Ratchet transport of an assembly of active run-and-tumble disks was realized in a traveling-wave substrate \cite{Sandor}. The collection of bacteria is able to migrate against the funnel-shaped barriers by creating and maintaining a chemoattractant gradient \cite{Lambert}. Li and coworkers \cite{Li} manipulated the transport of the overdampd point-like Janus particles in narrow two-dimensional corrugated channels. The chiral active particles can be rectified and sorted in complex and crowded environments \cite{Ai1,Mijalkov}. Theoretical and numerical studies show that the rectification effect is a general phenomenon that occurs for self-propelled particles in the presence of asymmetric periodic structures.

\indent  In most active ratchet systems the active particles are considered to be moving in the overdamped regime(at the low Reynolds number regime) \cite{Reichhardt0}.  However, the overdamped approximation is not justified in many situations (e. g. at the high Reynolds number regime) \cite{Nagai} such as self-propelling microdiodes \cite{Sharma}, Janus particles (microparticles) moving through a dusty plasm \cite{Ivlev} (air, or even a vacuum), colloidal particles in air, granular matter in dilute systems, and so on. In these systems, the damping is significantly reduced so that inertial effects can play an important role.  As we known, in non-active systems the inertia of the particle could exhibit peculiar behaviors \cite{Jung,Lind,Mateos}, such as chaotic transport, current reversals, and so on. Rectified transport of the active particles with underdamped dynamics remains largely unexplored. Therefore, it would be interesting to study the ratchet transport of active particles in the underdamped case. In this paper, we numerically study the minimal model of active ratchet with underdamped dynamics. We focus on finding how the inertia influences rectification of self-propelled particles. It is found that due to the existence of the inertial term  current reversals can occur by changing the friction coefficient or the self-propulsion force.

\section{Model and methods}
\begin{figure}[htpb]
\vspace{1cm}
  \label{fig:asym_contour}\includegraphics[width=0.6\columnwidth]{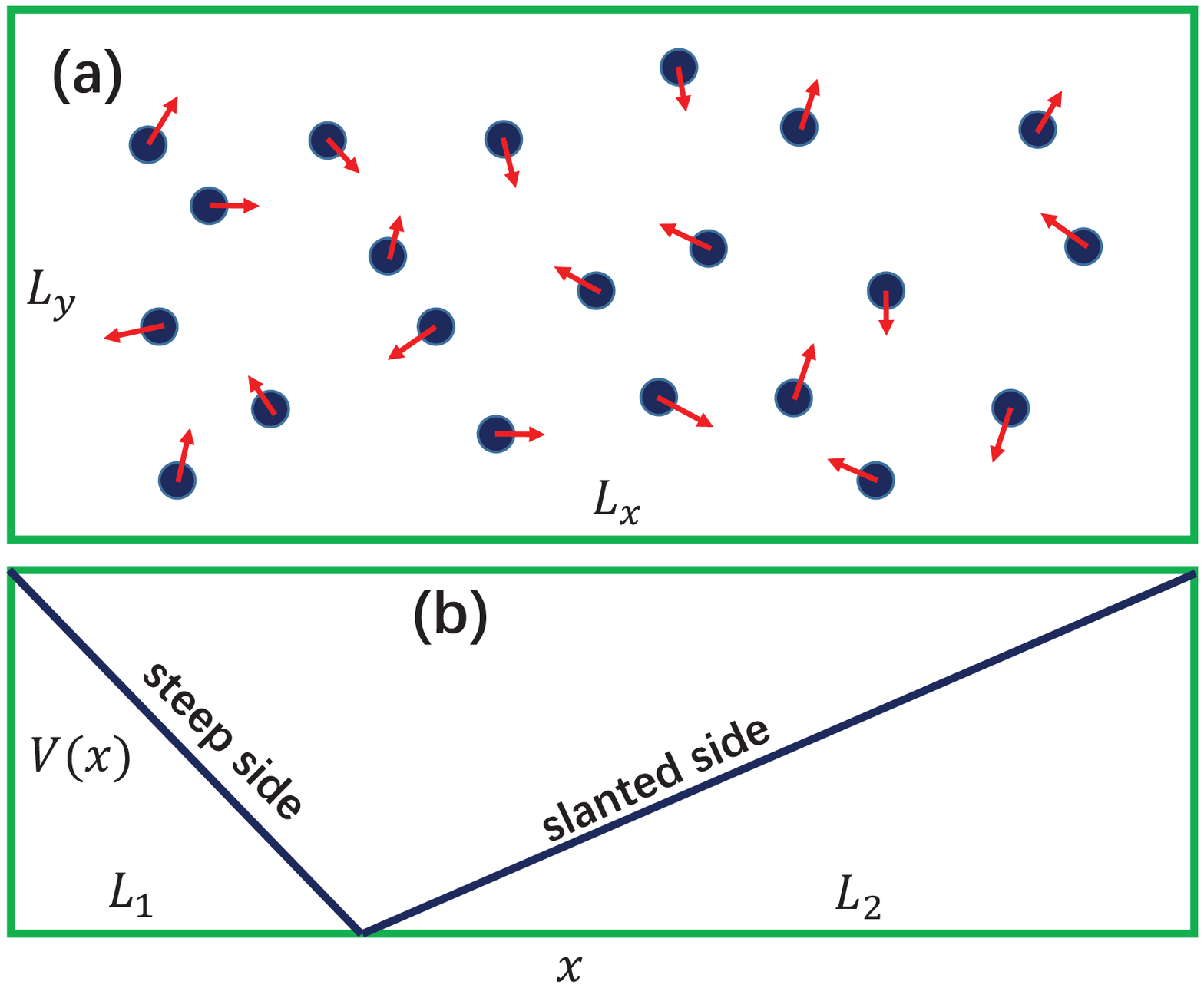}
  \caption{(Color online)(a) Schematic of the ratchet device: noninteracting point self-propelled particles moving in a straight periodic channel with width $L_y$ and period $L_x$. Periodic boundary conditions imposed in the $x$-direction and hard wall boundary condtions in the $y$-direction. (b) The profile of $x$-direction asymmetric potential $V(x)$ described in Eq.(4) at $\Delta=-\frac{1}{3}$, where the substrate force $f_p(x)$ is equal to $3.0$ for the steep side and $-1.5$ for the slanted side.
    \label{fig:Parallel_pressure_current}}
\end{figure}

\indent We consider noninteracting self-propelled particles with radius $r$ moving in a two-dimensional straight channel with confining walls at $y=0$ and $y=L_y$ as showed in Fig. 1(a).  In the channel, particles experience a substrate potential $V(x)=V(x+L_x)$ with period $L_x$ along $x$-direction.  In order to concentrate on the inertial effects on the ratchet transport, the interactions between particles are neglected.  The dynamics of each particle are described by the position $\mathbf{r}\equiv(x,y)$ of its center and the orientation $\theta$ of the polar axis $\mathbf{n}\equiv(\cos\theta,\sin\theta)$.  The particle obeys the following Langevin equations

\begin{equation}\label{e1}
  m\frac{d^2 x}{dt^2}=-\Gamma\frac{dx}{dt}+f_0\cos\theta+f_p(x)+\sqrt{2D_0}\zeta_x(t),
\end{equation}

\begin{equation}\label{e2}
   m\frac{d^2 y}{dt^2} =-\Gamma\frac{dy}{dt}+f_0\sin\theta+\sqrt{2D_0}\zeta_y(t),
\end{equation}
\begin{equation}\label{e3}
  \frac{d \theta}{dt}=\sqrt{2D_{\theta}}\zeta_\theta(t),
\end{equation}
where $m$ is the mass of the particle and $f_0$ is the self-propulsion force. $\Gamma$ is the Stokes friction coefficient. For small values of $\Gamma$ (e. g. particles in a gas or a dilute system), the interactions from the medium can be neglected and the inertia dominates the transport.  In the limit of large $\Gamma$ (e. g. particles in a liquid), the inertia of the particle can be neglected, the system reduces to its overdamped version. $D_0$ and $D_{\theta}$ denote the translational and rotational diffusion, respectively. $\zeta_{x,y,\theta}(t)$ is zero mean white noise with the relation $\langle \zeta_{i}(t)\xi_{j}(s)\rangle = \delta_{ij}\delta(t-s)$, where $i,j=x, y, \theta$. The symbol $\langle...\rangle$
denotes an ensemble average over the distribution of the random forces. $\delta$ is the Dirac delta function.

\indent $f_p(x)$ is the substrate force along $x$-direction which arises from an asymmetric potential $V(x)$ shown in Fig. 1(b). The profile of the potential within one period $0\leq x\leq L_x$ is described by
\begin{equation}\label{Vx}
     V(x)=
\begin{cases}
\frac{V_0}{L_1}(L_1-x), & \text{$0\leq x<L_1$};\\
\frac{V_0}{L_2}(x-L_1), & \text{$L_1\leq x\leq L_x$},
\end{cases}
\end{equation}
where $V_0$ is the height of the potential. We define the asymmetric parameter of the potential $\Delta=L_1-L_2$ and the potential is completely symmetric at $\Delta=0$.

\indent Eqs.(\ref{e1},\ref{e2},\ref{e3}) can be rewritten in the dimensionless forms by introducing the characteristic length scale the time scale: $\hat{x}=\frac{x}{L_x}$,$\hat{y}=\frac{y}{L_x}$, $\hat{t}=\frac{t}{\tau_0}$, and $\tau_0^2=\frac{m L_x^2}{V_0}$,
\begin{equation}\label{e4}
  \frac{d^2 \hat{x}}{d\hat{t}^2}=-\hat{\Gamma}\frac{d\hat{x}}{d\hat{t}}+\hat{f}_0\cos\hat{\theta}+\hat{f}_p(\hat{x})+\sqrt{2\hat{D}_0}\hat{\zeta}_{\hat{x}}(\hat{t}),
\end{equation}

\begin{equation}\label{e5}
  \frac{d^2 \hat{y}}{d\hat{t}^2}=-\hat{\Gamma}\frac{d\hat{y}}{d\hat{t}}+\hat{f}_0\sin\hat{\theta}+\sqrt{2\hat{D}_0}\hat{\zeta}_{\hat{y}}(\hat{t}),
\end{equation}
\begin{equation}\label{e6}
  \frac{d \hat{\theta}}{d\hat{t}}=\sqrt{2\hat{D}_{\theta}}\hat{\zeta}(\hat{t}),
\end{equation}
where $\hat{\Gamma}=\frac{\tau_0}{m}\Gamma$, $\hat{f}_0=\frac{f_0L_x}{V_0}$, $\hat{L}_y=\frac{L_y}{L_x}$, $\hat{D}_\theta=D_\theta\tau_0$, and $\hat{D}_0=\frac{D_0\tau_0}{mV_0}$. The potential is rewritten as
\begin{equation}\label{Vx1}
     \hat{V}(\hat{x})=
\begin{cases}
1-\frac{\hat{x}}{\hat{L}_1}, & \text{$0\leq \hat{x}<\hat{L}_1$};\\
\frac{1}{\hat{L}_2}(\hat{x}-\hat{L}_1), & \text{$\hat{L}_1\leq \hat{x}\leq 1$},
\end{cases}
\end{equation}
where $\hat{L}_1=\frac{L_1}{L_x}$ and $\hat{L}_2=\frac{L_2}{L_x}$. From now on, we will use only the dimensionless variables and shall omit the hat for all quantities occurring in Eqs. (\ref{e4},\ref{e5},\ref{e6},\ref{Vx1}). One can obtain the substrate force in dimensionless form $f_p(x)=\frac{2}{1+\Delta}$ for $0\leq x<\frac{1+\Delta}{2}$ and $\frac{2}{\Delta-1}$ for $\frac{1+\Delta}{2}\leq x\leq 1$.

\indent When the inertia of the particle is ignored (the overdamped case), Eqs. (\ref{e4},\ref{e5}) can be rewritten as
\begin{equation}\label{e7}
 {\Gamma}\frac{d{x}}{d{t}}={f}_0\cos{\theta}+{f}_p(x)+\sqrt{2D_0}\zeta_x(t),
\end{equation}

\begin{equation}\label{e8}
  {\Gamma}\frac{d{y}}{d{t}}={f}_0\sin{\theta}+\sqrt{2D_0}\zeta_y(t).
\end{equation}

\indent Because particles are confined in the $y$-direction, the directed transport only occurs in the $x$-direction. To quantify the ratchet effect, we measure the average velocity in the $x$-direction. In the asymptotic long-time regime, the average velocity of the particle along the $x$-direction can be obtained from the following formula
\begin{equation}\label{velocty}
  V_{\theta_0}=\lim_{t\rightarrow\infty}\frac{\langle x(t)\rangle_{\theta_0}}{t},
\end{equation}
where $\theta_0$ is the initial angle of the trajectory. The full average velocity after a second average over all $\theta_0$  is $V_s=\frac{1}{2\pi}\int_{0}^{2\pi} V_{\theta_0}d\theta_0$.

\indent Periodic boundary conditions and hard wall boundary conditions are imposed in the $x$ and $y$-directions, respectively. The particle-wall interactions was modeled as follows. When the particle meets the wall, its translational position is elastically reflected. The rotation of the angle $\theta$ is induced by a tangential friction and the angle is assumed to be randomized. The behavior of the quantities of interest can be corroborated by integration of the Langevin Eqs.(\ref{e4},\ref{e5},\ref{e6}) using the second-order stochastic Runge-Kutta algorithm. In our simulations, the integration step time $\delta t$ was chosen to be smaller than $10^{-3}$ and the total integration time was more than $10^7$. The stochastic averages reported above were obtained as ensemble averages over $10^4$ trajectories with random initial conditions.
\section{Results and Discussion}

\subsection{Zero translational diffusion case}
\indent We firstly consider the zero translational diffusion case $D_0=0.0$ with $\Delta=-\frac{1}{3}$ and $L_y=2.0$. For this case (showed in Fig. 1(b)), the substrate force $f_p$ is equal to $3.0$ for the steep side and $-1.5$ for the slanted side. In order to facilitate discussion, we introduce the effective driving force along $x$-direction $F_d(t)=-\Gamma\frac{dx}{dt}+f_0\cos\theta(t)$ for the underdamped case and $F_d(t)=f_0\cos\theta(t)$ for the overdamped case. The probability distribution $\rho(f_d)$ for $F_d(t)$ is a useful tool for investigation of the ratchet transport. We define $\rho(f_d)$ as a function that gives the probability that $F_d(t)$ is exactly equal to the value $f_d$. We vary $f_0$, $\Gamma$, and $D_\theta$ and measure the resulting average velocity of the ratcheting behavior.

\begin{figure}[htpb]
\vspace{1cm}
  \label{fig1}\includegraphics[width=0.45\columnwidth]{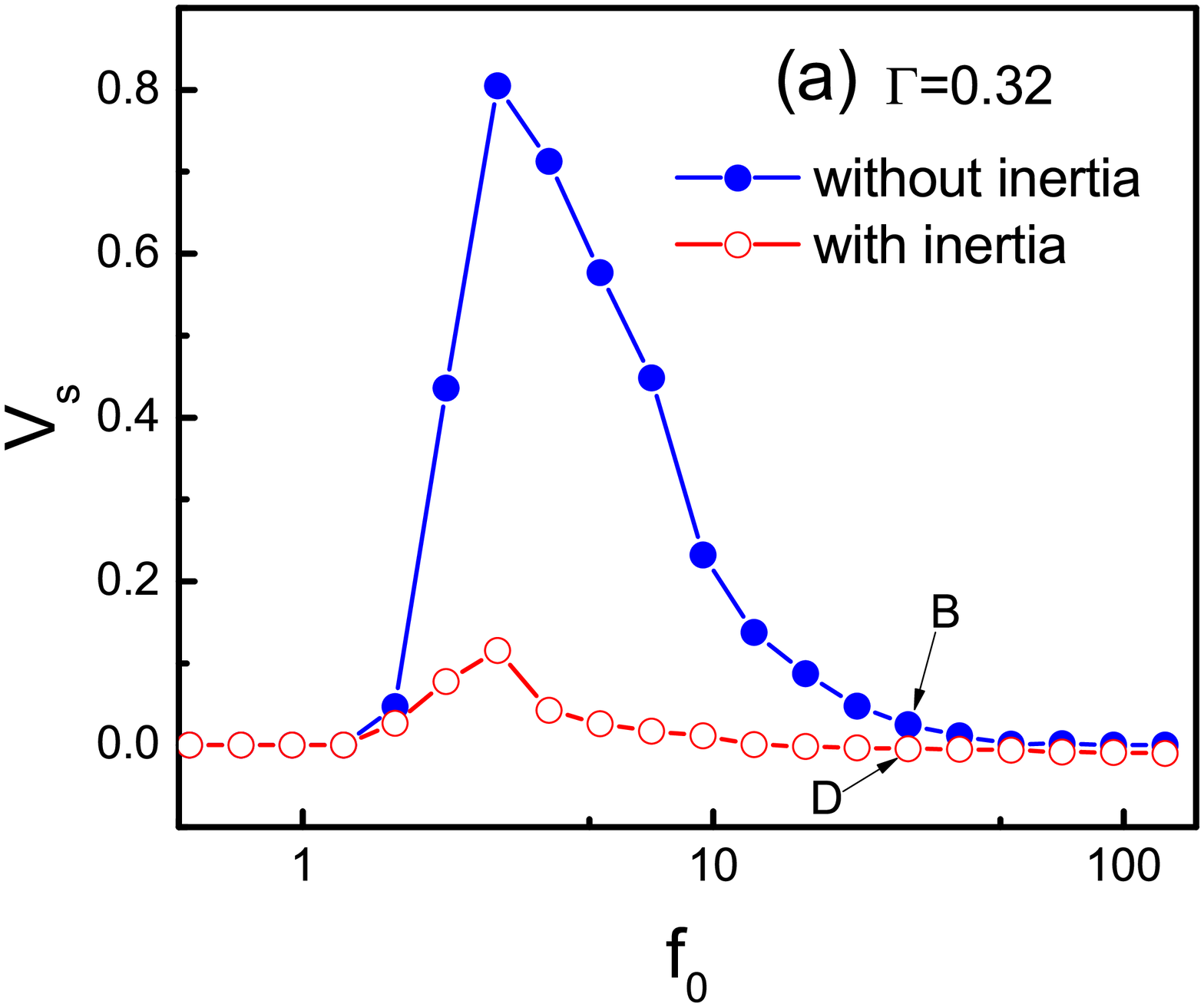}
  \includegraphics[width=0.45\columnwidth]{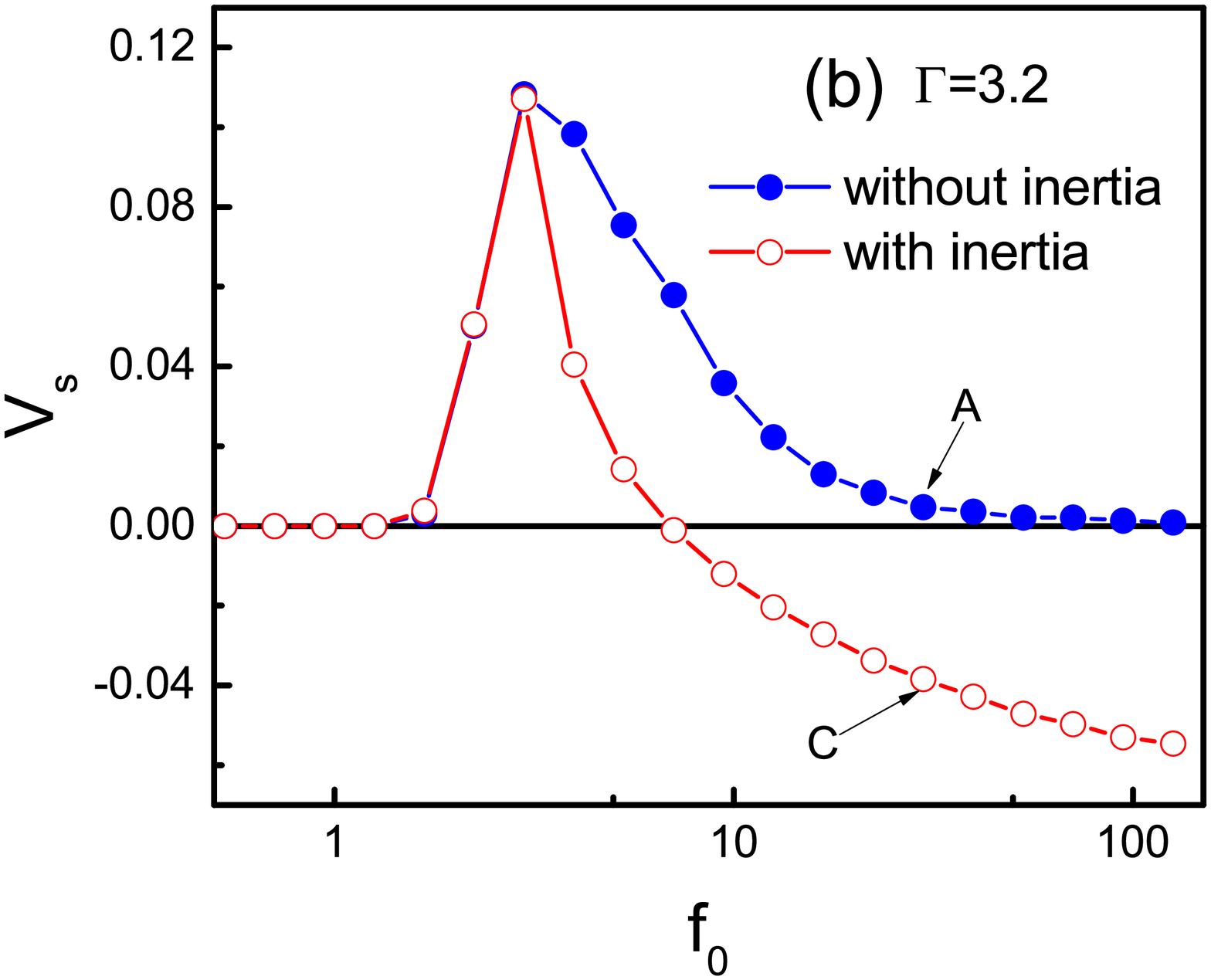}
  \includegraphics[width=0.45\columnwidth]{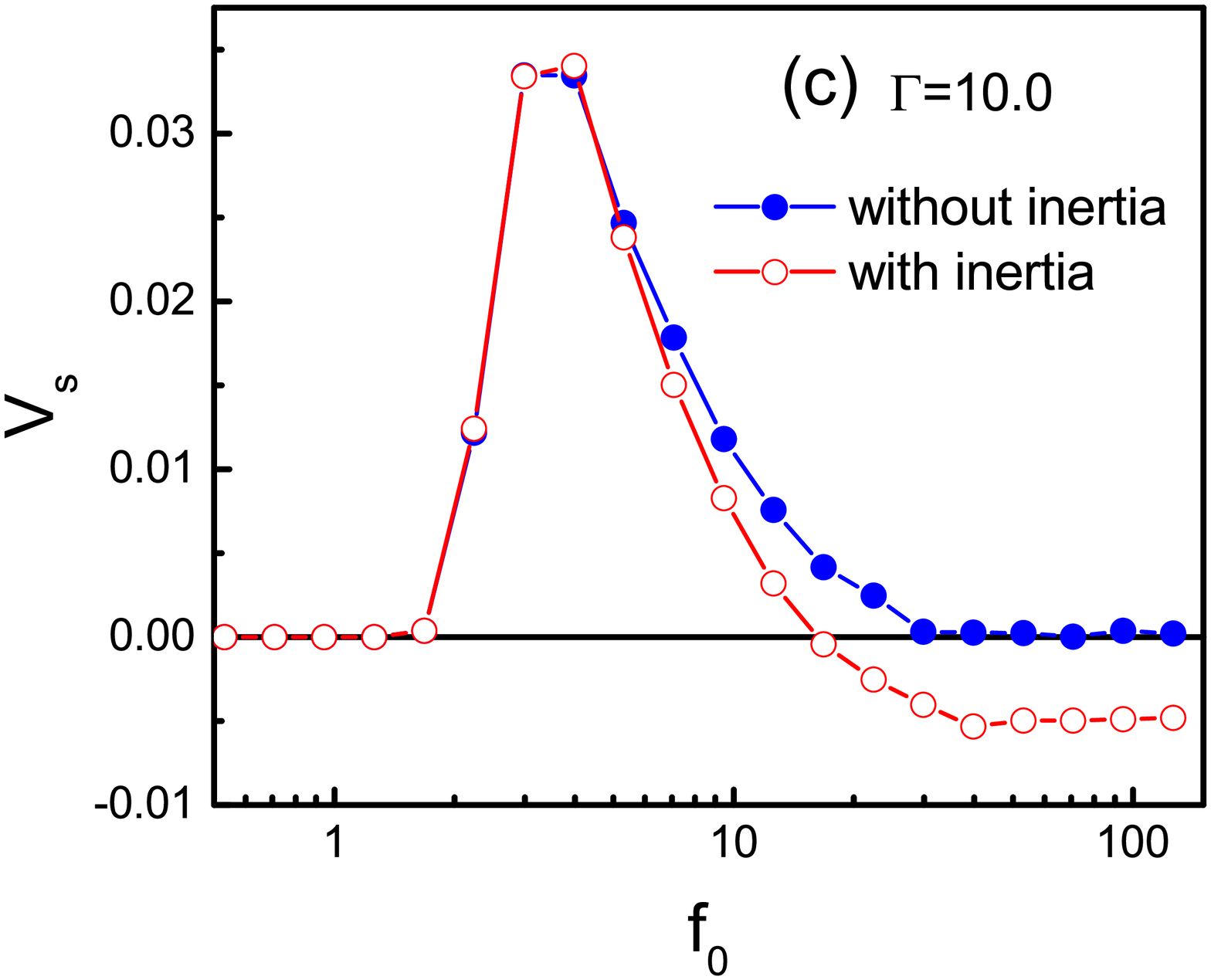}
  \includegraphics[width=0.45\columnwidth]{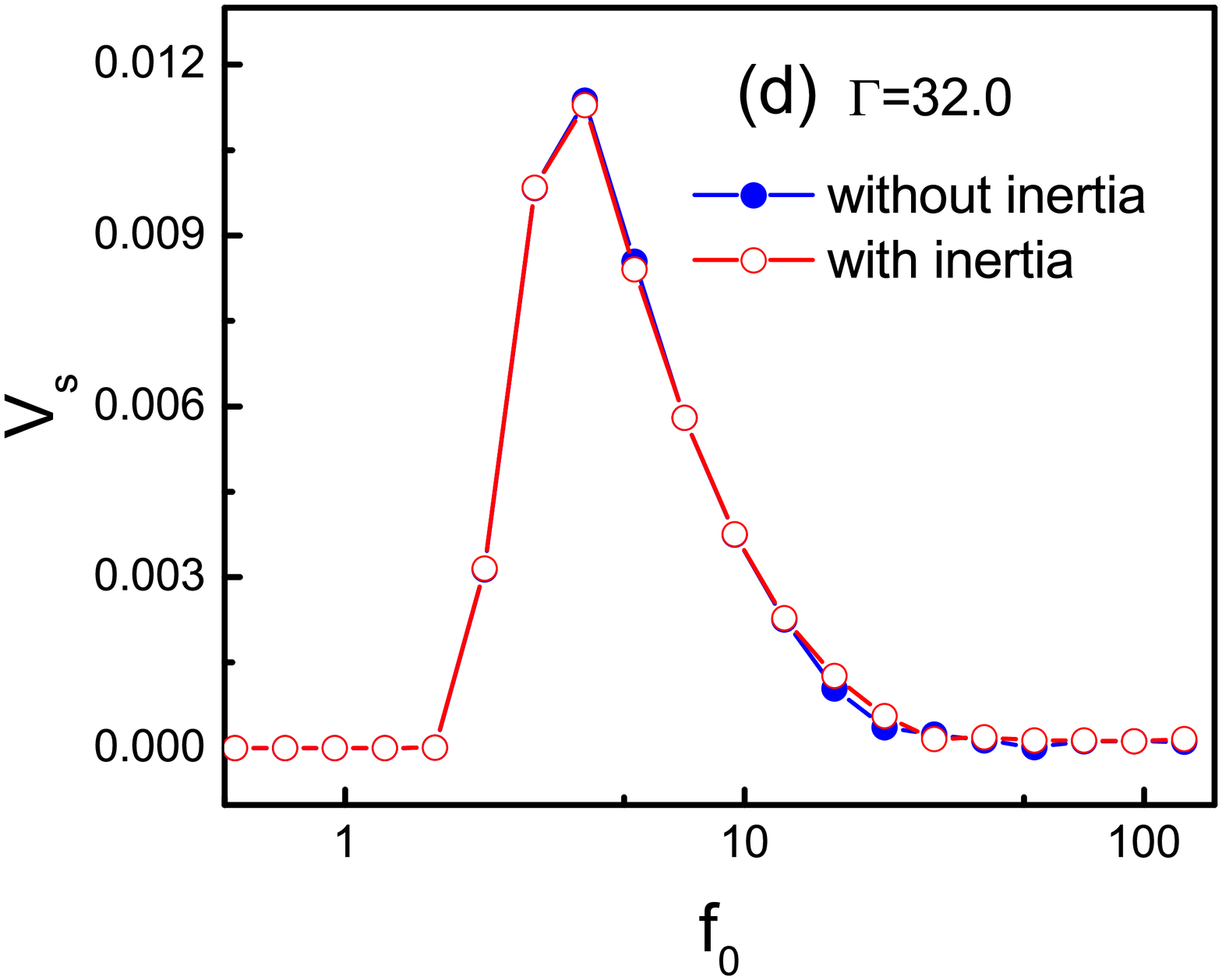}
  \caption{(a) (Color online)Average velocity $V_s$ as a function of the self-propulsion force $f_0$ at $D_\theta=0.01$ when the inertial term is included as well as neglected.
 }
\end{figure}
\indent Figure 2 describes the average velocity $V_s$ as a function of the self-propulsion force $f_0$ when the inertial term is included as well as neglected. For the overdamped case (without inertial term), the particle always moves along the easy direction of the substrate asymmetry (positive $V_s$) and no current reversals occur. The ratchet phenomenon can be explained as follows.  When $\Delta=-\frac{1}{3}$, the left side from the minima of the potential is steeper, it is easier for particle to move toward the slanted side than toward the steeper side, so the particle on average moves to the right(positive average velocity).  When $f_0<1.5$ and $|f_d(t)|$ is always less than $1.5$, the particle cannot pass across the barrier and stays in the potential well, thus the average velocity is equal to zero.  When $f_0\gg 3$, the effective driving force $|f_d(t)|$ is much larger than the substrate force $|f_p|$, the substrate potential can be negligible, thus the ratchet effect disappears and $V_s$ tends to zero. Therefore, there exists an optimal value of $f_0$ (about $f_0=3.0$) at which the average velocity takes its maximal value. It is important to note that in the overdamped limit current reversals of non-active particles can be induced by several combined rectification mechanisms in the complex systems \cite{Malgaretti,Reimann}.  However, in our simple system,  the ratchet effect only occurs in a single direction when the inertia is neglected.

\indent For the underdamped case (with inertial term), the transport behaviors become complex. For very small value of $\Gamma$ (e. g. $\Gamma=0.32$), the average velocity is always positive.  For very large value of $\Gamma$ (e. g. $\Gamma=32$) showed in Fig. 2(d), the system reduces to its overdamped version, the curves are completely identical for both underdamped and overdamped cases. However, for intermediate values of $\Gamma$ (e. g. $\Gamma=3.2$ and $10$) showed in Figs. 2(b) and 2(c), the curves for the underdamped case are different from those for the overdamped case. For the underdmaped case, on increasing $f_0$, the average velocity firstly increases to its maximal value, then decreases, and finally reverses its direction. The most important feature for the underdamped case is the occurrence of current reversals for the intermediate values of $\Gamma$. Note that current reversal is also a common feature in collectively interacting non-active systems\cite{Silva,Lu,Lara,Reichhardt3}, where the reversal mechanism is different from that in the present system.

\begin{figure}[htpb]
  \label{fig3}\includegraphics[width=0.45\columnwidth]{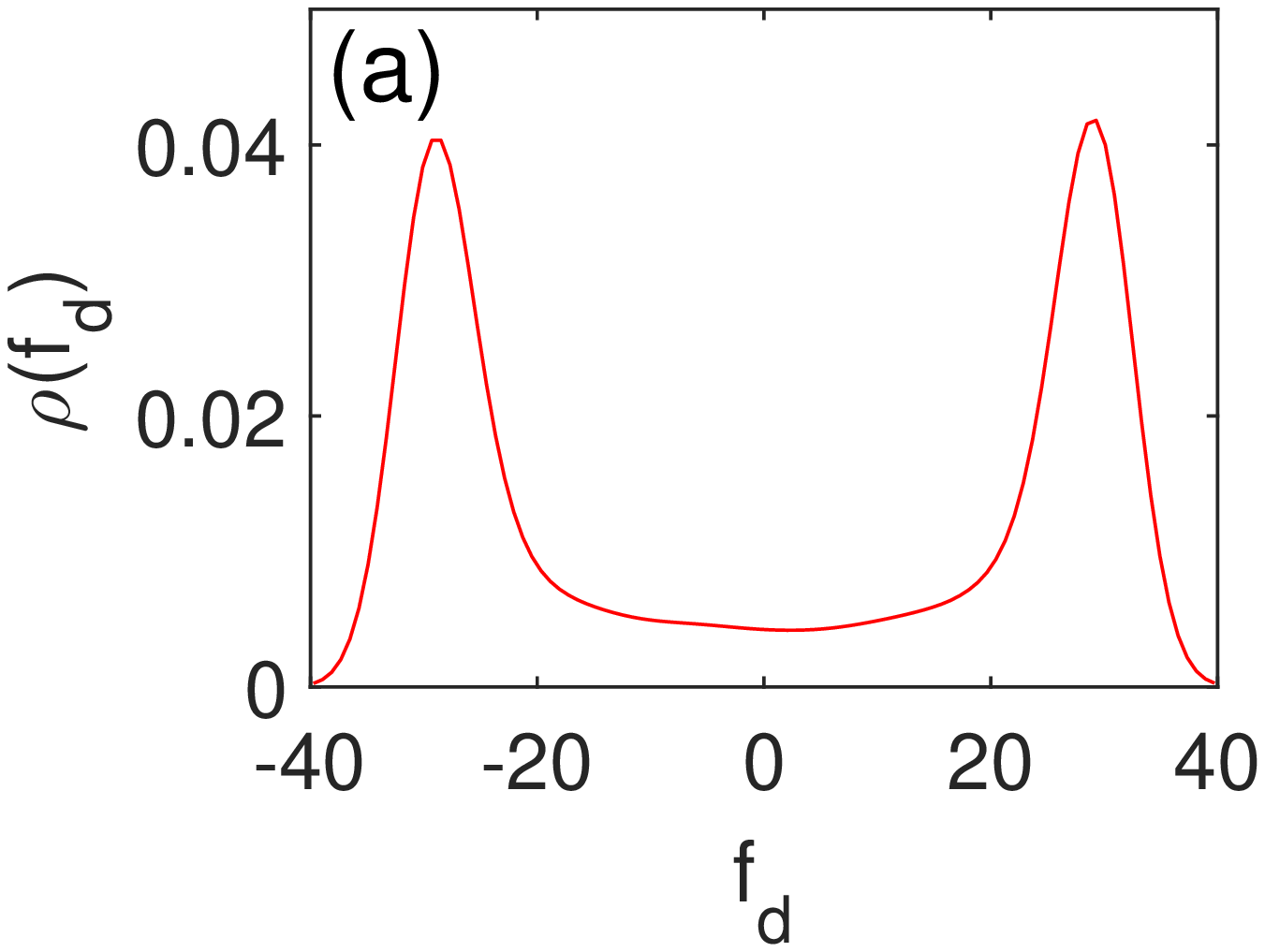}
  \includegraphics[width=0.45\columnwidth]{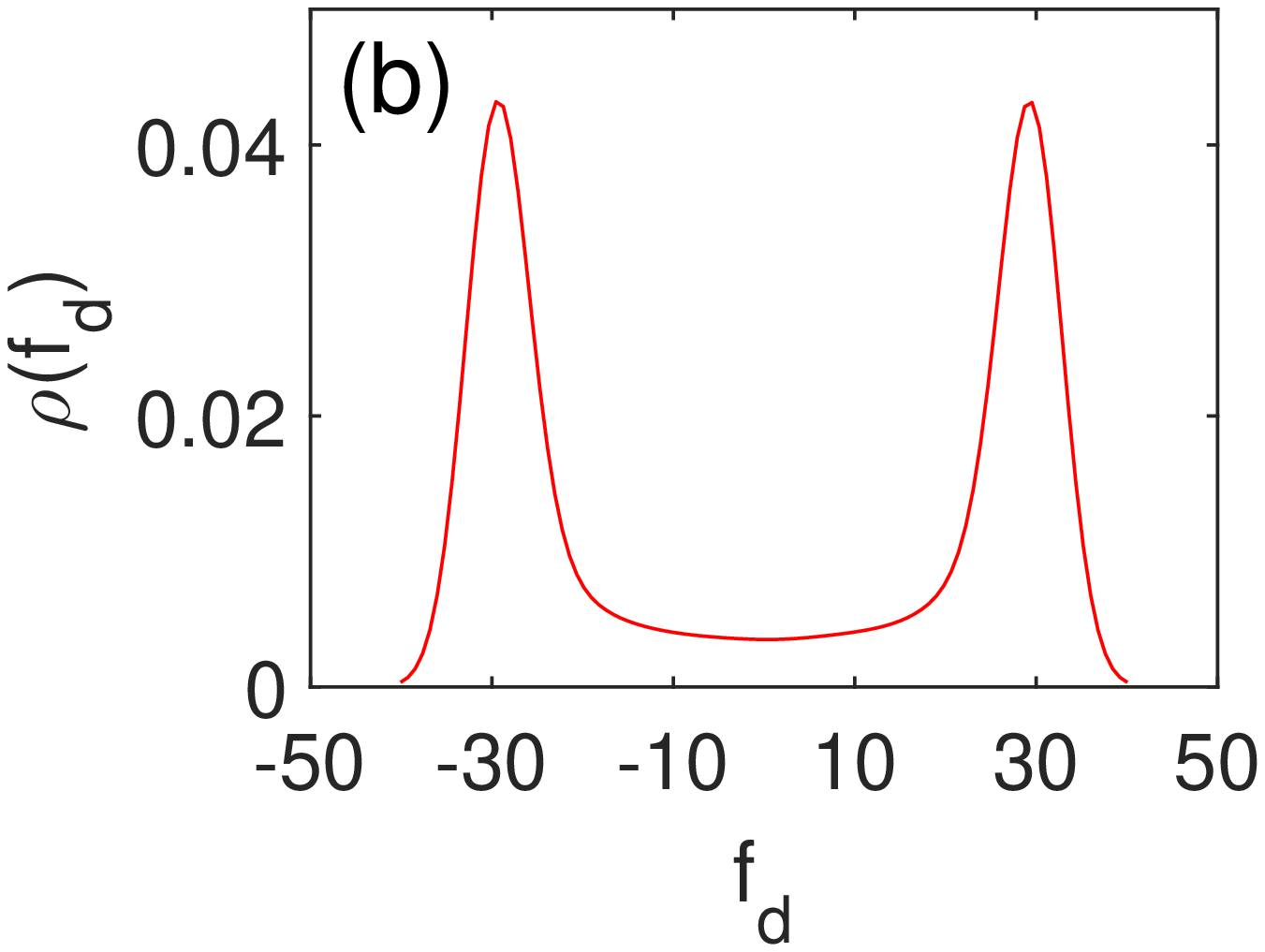}
  \includegraphics[width=0.45\columnwidth]{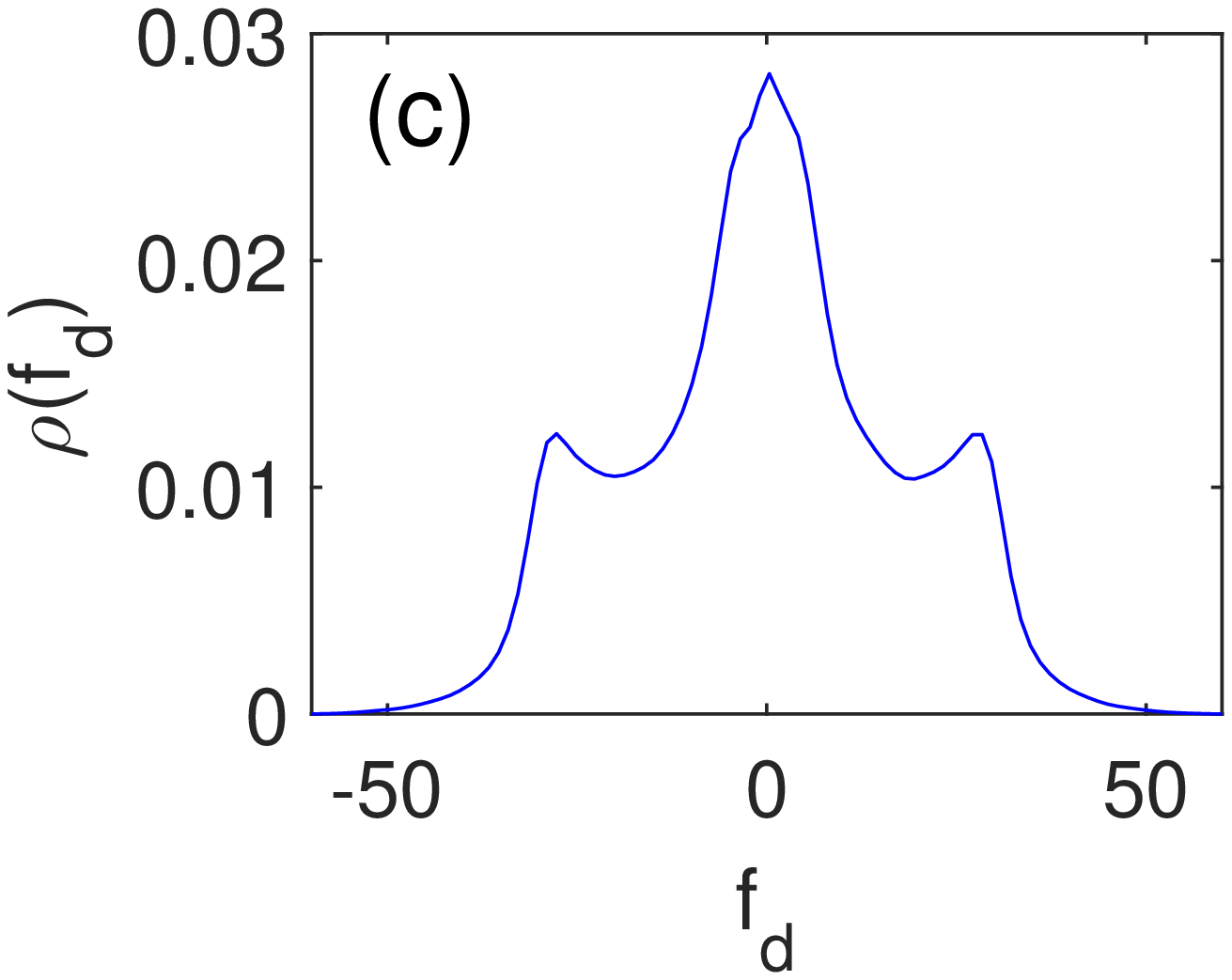}
  \includegraphics[width=0.45\columnwidth]{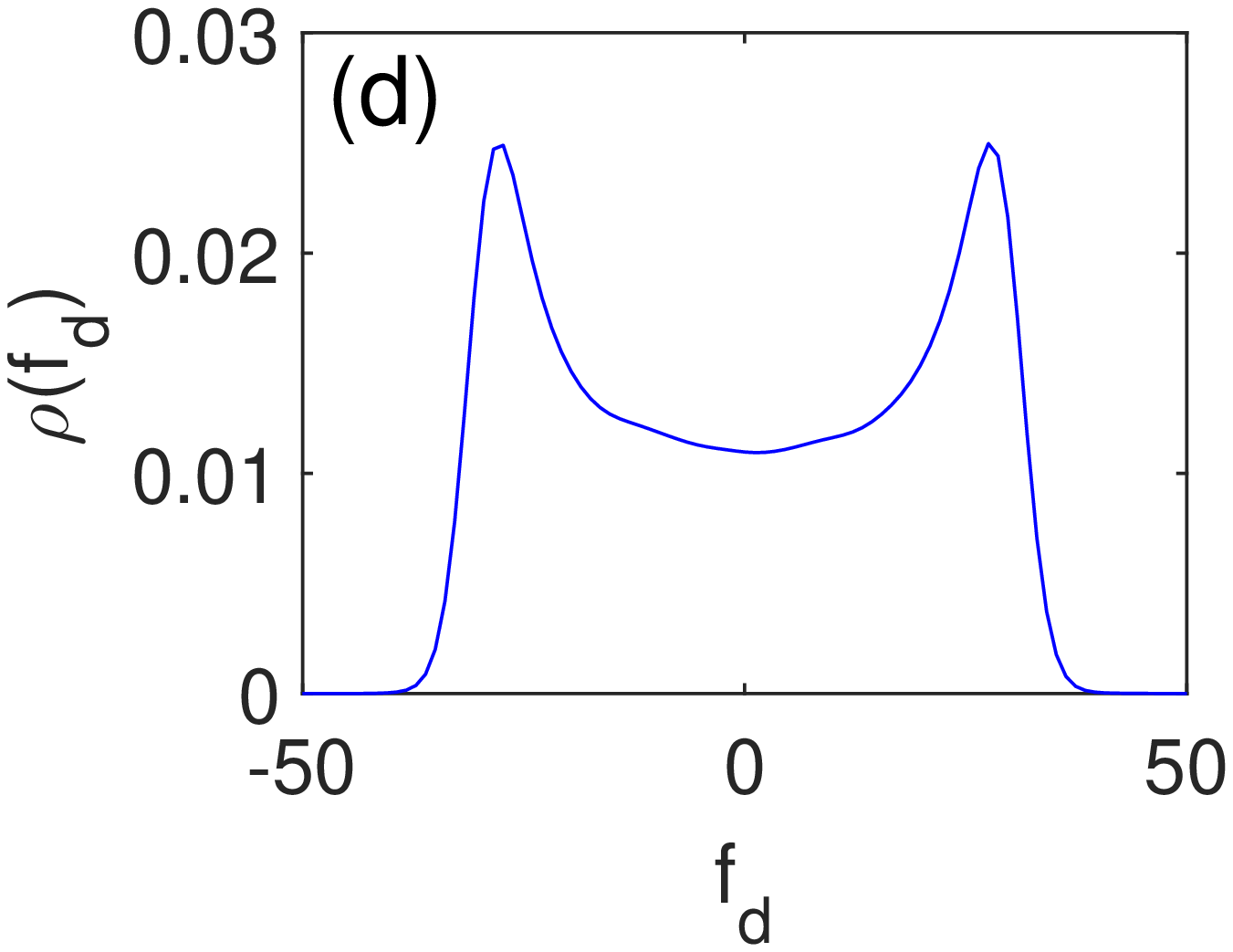}
  \caption{Distribution $\rho(f_d)$ of the effective driving force at $f_0=30.0$. (a)Overdamped case at $\Gamma=3.2$. (b)Overdamped case at $\Gamma=0.32$. (c)Underdamped case at $\Gamma=3.2$. (d)Underdamped case at $\Gamma=0.32$.
 }
\end{figure}

\indent Now we will give the detailed explanation for the above behaviors. The direction of transport is determined by the competition between the slope of the potential and the distance from minima to maxima of the potential. When the slope of the potential dominates the transport (particles cannot easily pass across the barrier), the left side from the minima of the potential is steeper, thus particle on average moves to the right. When the distance from minima to maxima of the potential dominates the transport (particles can easily pass across the barrier), the distance from the slanted side is larger than that from the steep side, thus particle is easily thrown out from the steep side, resulting in negative velocity. Of course, if the potential can be  negligible, the ratchet effect will disappear.

\indent We will discuss four special points in Fig. 2. The distributions $\rho(f_d)$ for points $A$, $B$, $C$, and $D$ in Figs. 2(a) and 2(b) are respectively shown in Figs. 3(a)-3(d). For the overdamped case (points $A$ and $B$), the effective driving force is mainly distributed at $f_d=-30$ and $30$ (two peaks at $-30$ and $30$),  the effective driving force is much larger than the substrate force, the substrate potential can be negligible, thus the ratchet effect disappears and $V_s$ tends to zero.
However, for the underdamped case, due to the existence of the term $-\Gamma\frac{dx}{dt}$, the distribution is different from that for the overdamped case.  For point $C$, the effective driving force is mainly distributed at $f_d=-30$, $0$, and $30$ shown in Fig. 3(c). Though some values of $F_d(t)$ are much larger than the substrate force, the role of the substrate potential cannot be negligible (due to the existence of the peak near $f_d=0$). In this case, the distance from minima to maxima of the potential dominates the transport, which results in a negative velocity. For point $D$, due to the very small values of $\Gamma$ (e. g. 0.32), the effective driving force is mainly distributed at $f_d=-30$ and $30$ shown in Fig. 3(d), which is similar to the overdamped case (e.g. points $A$ and $B$). Therefore, in this case, the substrate potential can be negligible and the ratchet effect disappears.

\begin{figure}[htpb]
\vspace{1cm}
  \label{fig1}\includegraphics[width=0.8\columnwidth]{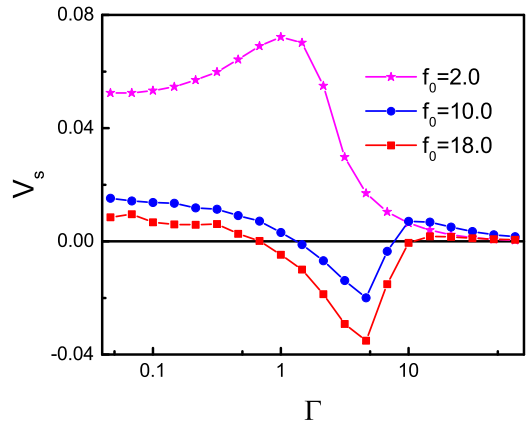}
  \caption{(a) (Color online)Average velocity $V_s$ as a function of the friction coefficient $\Gamma$ for different values of $f_0$ at $D_\theta=0.01$.}
\end{figure}

\indent Figure 4 displays the average velocity $V_s$ as a function of the friction coefficient $\Gamma$.  When $f_0<3.0$ (e. g. $f_0=2.0$),  the slope of the potential dominates the transport. The average velocity is always positive and there exists an optimal value of $\Gamma$ at which $V_s$ is maximal.  When $f_0\gg 3.0 $ (e. g. $f_0=10.0, 18.0$), the average velocity changes its direction for serval times, which is due to the competition between the substrate force $f_p$ and the effective driving force $F_d(t)$. When $\Gamma\rightarrow \infty$, the self-propulsion force is completely suppressed, and thus the average velocity tends to zero. Therefore, we can have current reversals by changing the friction coefficient $\Gamma$.

\begin{figure}[htpb]
\vspace{1cm}
  \label{fig2}\includegraphics[width=0.8\columnwidth]{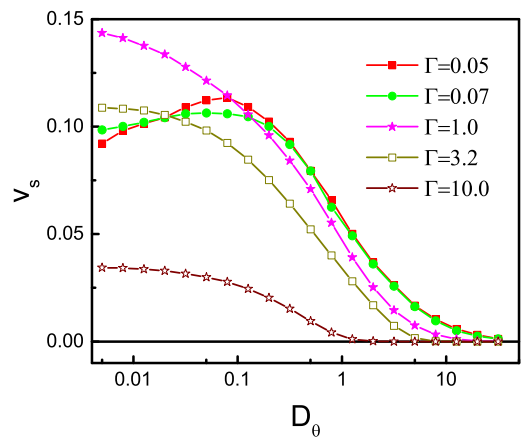}
  \caption{(a) (Color online)Average velocity $V_s$ as a function of the rotational diffusion coefficient $D_\theta$ for different values of $\Gamma$ at $f_0=3.0$.}
\end{figure}
\indent Figure 5 shows the average velocity $V_s$ as a function of the rotational diffusion coefficient $D_\theta$ for different values of $\Gamma$.
 The correlation function of the orientation is known as $\langle \mathbf{n}(t) \cdot\mathbf{n}(t^{'})\rangle=e^{-D_\theta |t-t^{'}|}$\cite{Schienbein}.
 When $D_{\theta}\rightarrow \infty$, this correlation reduces to a delta function and the self-propulsion force acts as a zero mean white noise. Thus, the system is in equilibrium and the average velocity tends to zero.  For small values of $D_\theta$, the transport behaviors are different for two cases: (I) for damping case (e. g. $\Gamma=1.0, 3.2, 10.0$) and (II)for very weak damping case (e. g. $\Gamma=0.05, 0.07$). For case I, the momentum relaxation time ($\tau_p=1/\Gamma$) is far less than the persistence time ($\tau_r=1/D_{\theta}=100$), the rotational diffusion dominates the transport. In this case, When $D_\theta\rightarrow 0$, the self-propelled angle almost does not change (the adiabatic regime), and the average velocity approaches its maximal value, which is similar to the adiabatic case in the forced thermal ratchet\cite{Magnasco}.  For case II, the momentum relaxation time $\tau_p$ becomes large, the rotational diffusion competes with the damping process. The increase of $D_\theta$ can cause two factors: (A)reducing the self-propelled driving, which blocks the ratchet transport, and (B) activating Brownian motion of particles, which facilitates the ratchet transport. When $D_\theta$ increases from zero, factor B dominates the transport and the average velocity is increased to its maximal value. When further increasing $D_\theta$,  factor A takes effect, the average velocity decreases. Therefore, for the damping case the average velocity decreases monotonously with increasing $D_\theta$ and for the very weak damping case there exists an optimal value of $D_\theta$ at which the average velocity $V_s$ is maximal.

\begin{figure}[htpb]
\vspace{1cm}
\centering
  \label{fig:asym_contour}\includegraphics[height=0.35\columnwidth]{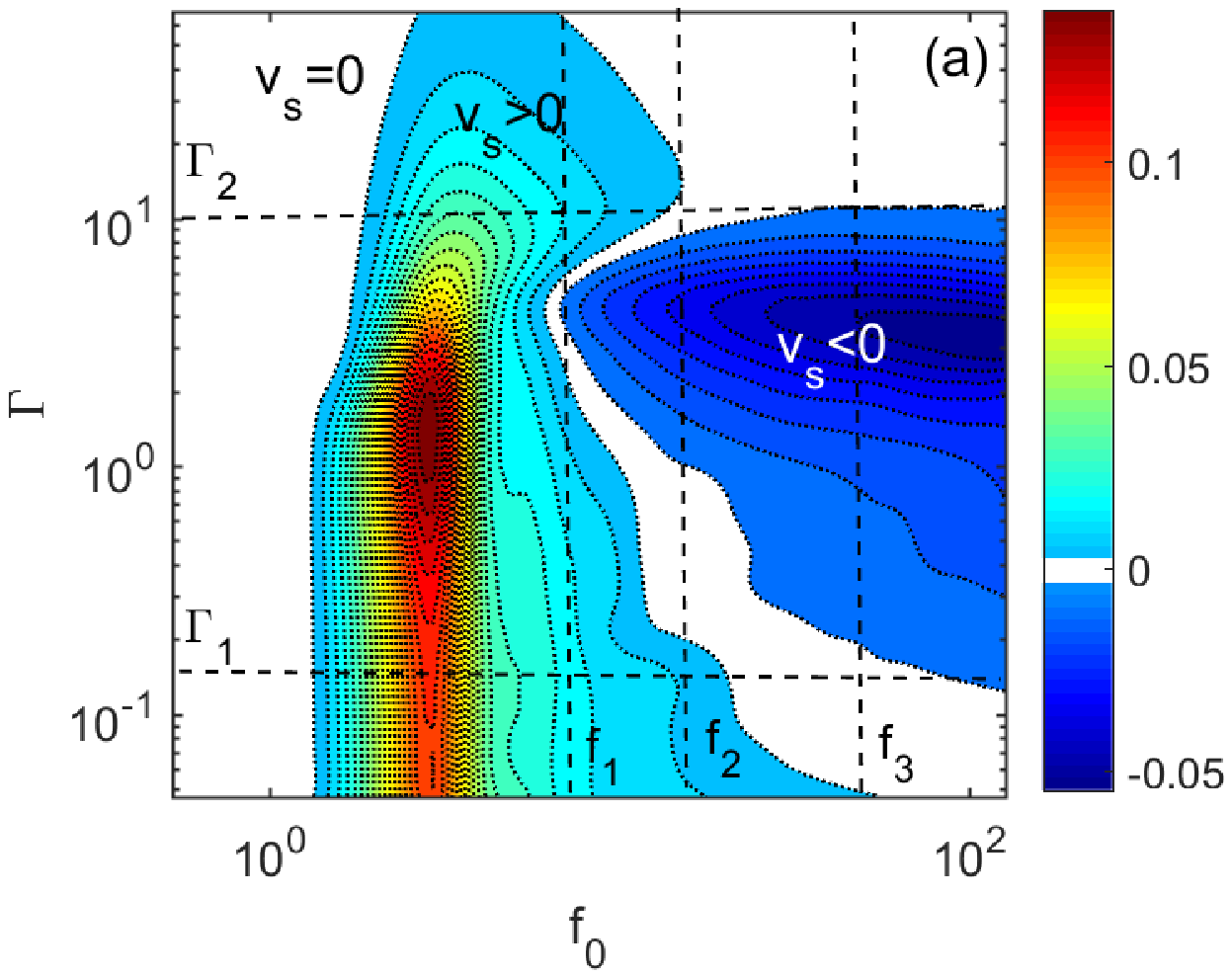}
  \includegraphics[height=0.35\columnwidth]{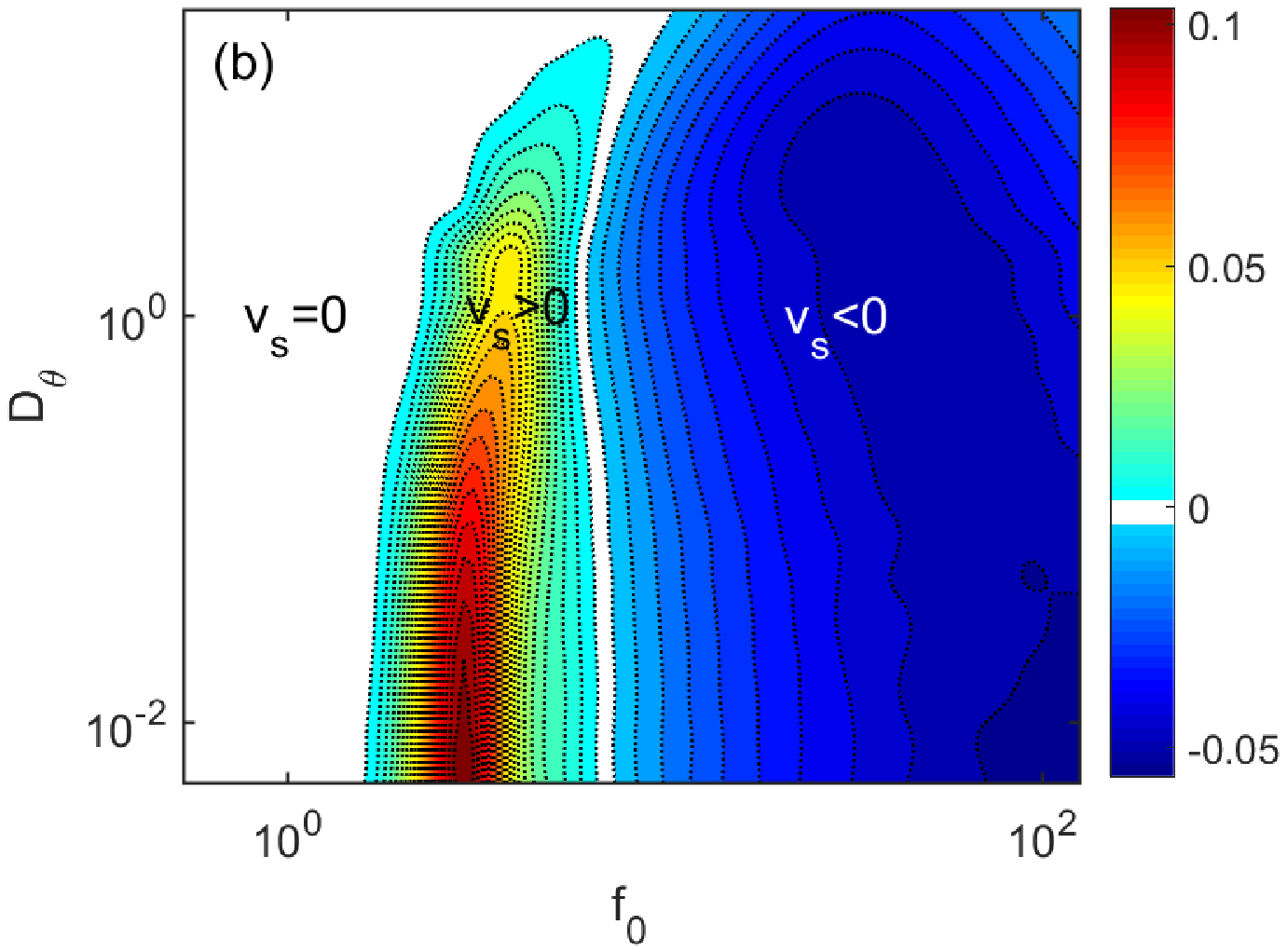}
  \caption{(a) (Color online)Contour plots of the average velocity $V_s$ as a function of the system parameters $\Gamma$ and $f_0$ at $D_\theta=0.01$. (b) Contour plots of the average velocity $V_s$ as a function of the system parameters $D_\theta$ and $f_0$ at $\Gamma=3.2$. The dotted lines are obtained from the numerical solution of the Langevin equations.
    \label{fig:Parallel_pressure_current}}
\end{figure}

\indent To study in more detail the dependence of rectification on $f_0$ and $\Gamma$, we calculated the average velocity as a function of $f_0$ and $\Gamma$ in Fig. 6(a).  When $\Gamma$ is increased from zero, the average velocity is always non-negative (non-positive) for $f_0<f_1$ ($f_0>f_3$), interestingly, it changes its direction twice for $f_1<f_0<f_2$ and one time for $f_2<f_0<f_3$. When $f_0$ is increased from zero, the average velocity is non-negative for $\Gamma<\Gamma_1$ and $\Gamma>\Gamma_2$ and it changes its direction when $\Gamma_1<\Gamma<\Gamma_2$. The critical values of $f_1$, $f_2$, $f_3$, $\Gamma_1$, and $\Gamma_2$ depend on the system parameters. Contour plots of the average velocity $V_s$ as a function of system parameters $D_\theta$ and $f_0$ at $\Gamma=3.2$ are shown in Fig. 6(b). It is found that the zero-value band (white band) between the positive region and the negative region is almost vertical, which means that on increasing $D_\theta$ from zero the net velocity almost does not change its direction.

\indent After the first demonstration of current reversals in the underdamped active particle system, it is soon realized that this response phenomenon is very sensitive to particle properties. This opens the possibility of steering different self-propulsion particles in opposite directions under identical experimental conditions. In our underdamped system, particles with the large self-propulsion force move to the left, whereas particles with the small self-propulsion force move to the right. Therefore, one can separate particles of different self-propulsion forces and make them move
towards opposite directions.

\subsection{Dependence of rectification on $\Delta$, $L_y$, and $D_0$}
\indent In order to present more comprehensive information of the rectification, we investigate the dependence of rectification on the parameters $\Delta$, $L_y$, and $D_0$ at $D_\theta=0.01$ and $\Gamma=3.2$.
\begin{figure}[htpb]
\vspace{1cm}
\centering
  \label{fig:asym_contour}\includegraphics[width=0.45\columnwidth]{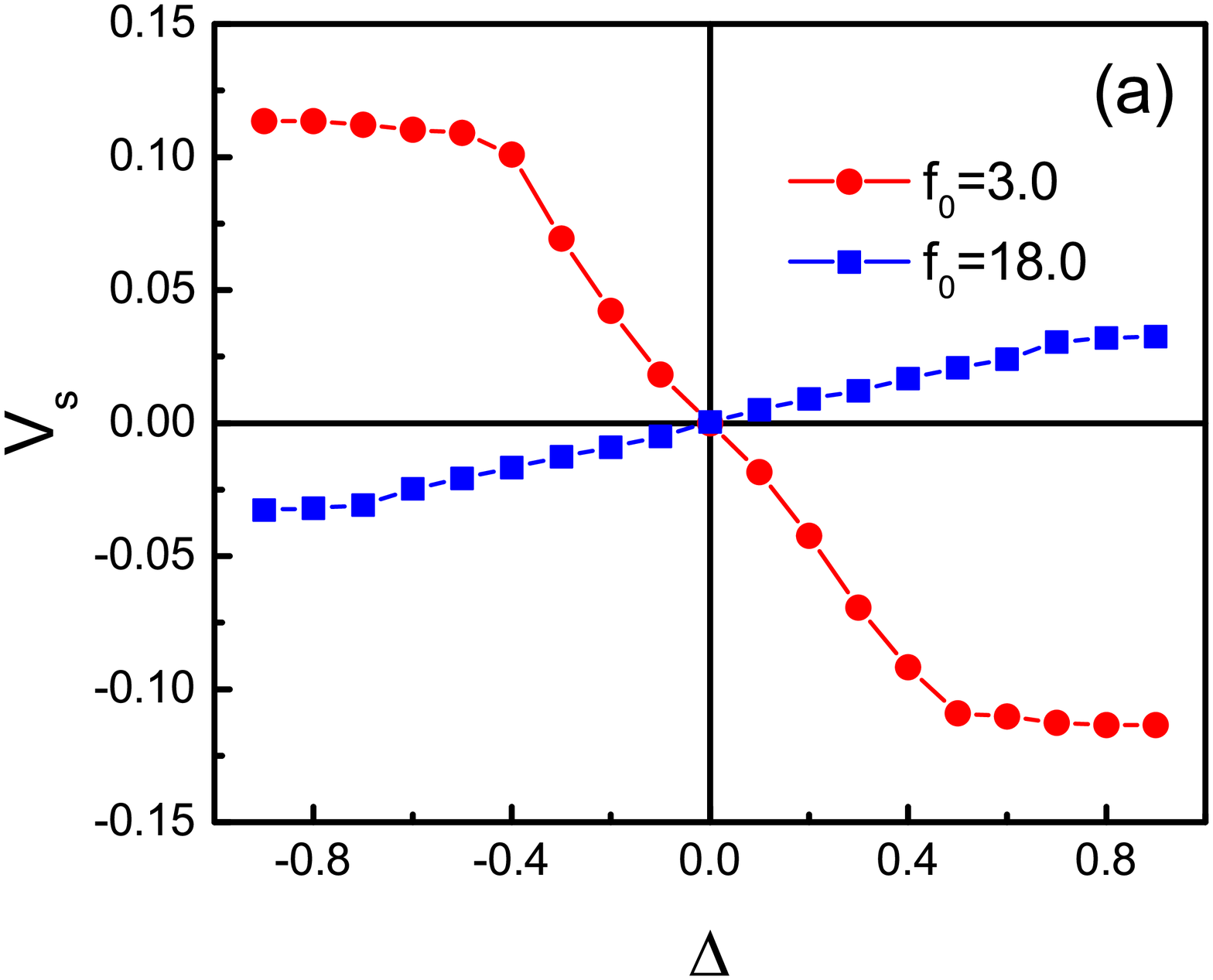}
  \includegraphics[width=0.45\columnwidth]{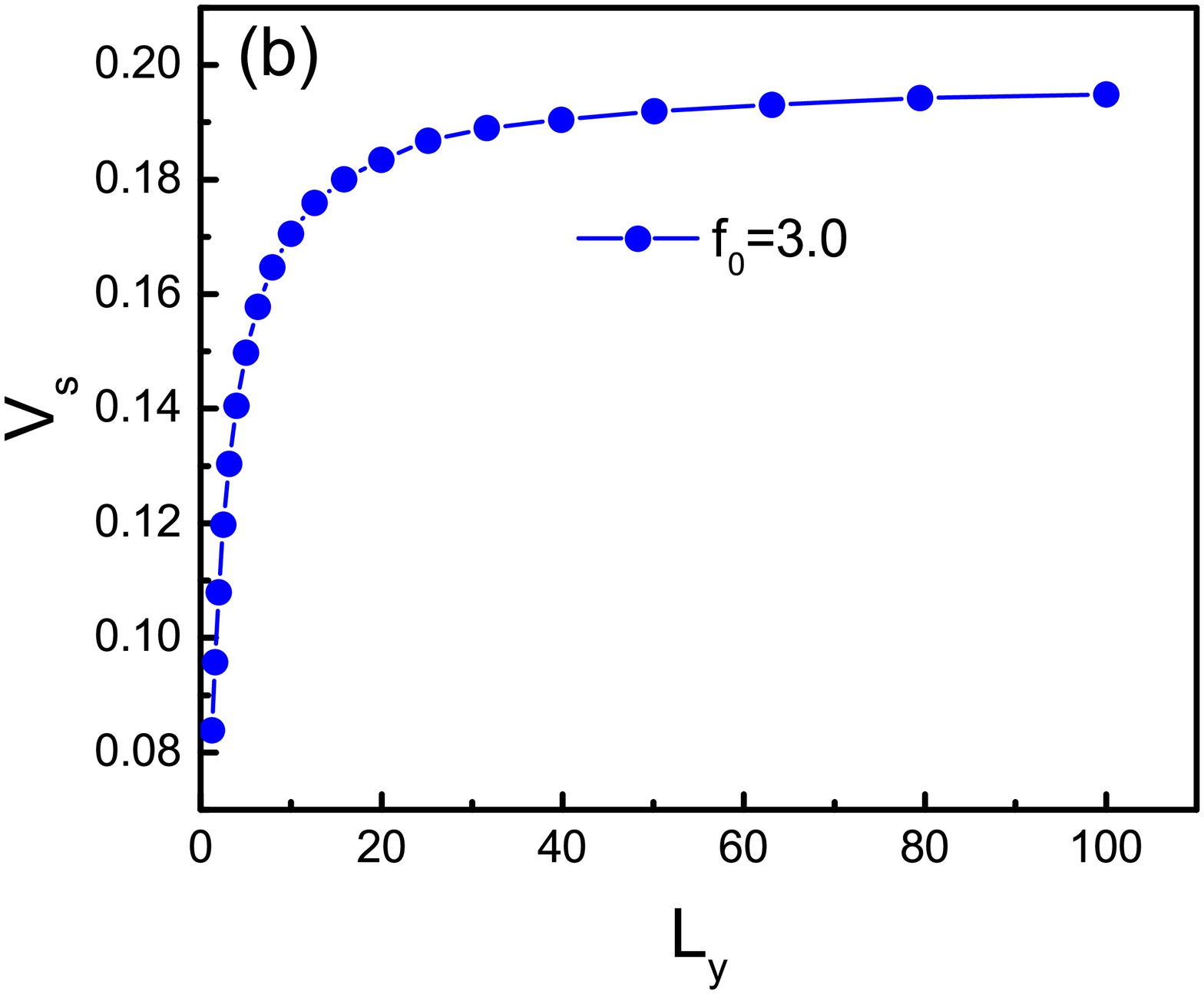}
  \includegraphics[width=0.45\columnwidth]{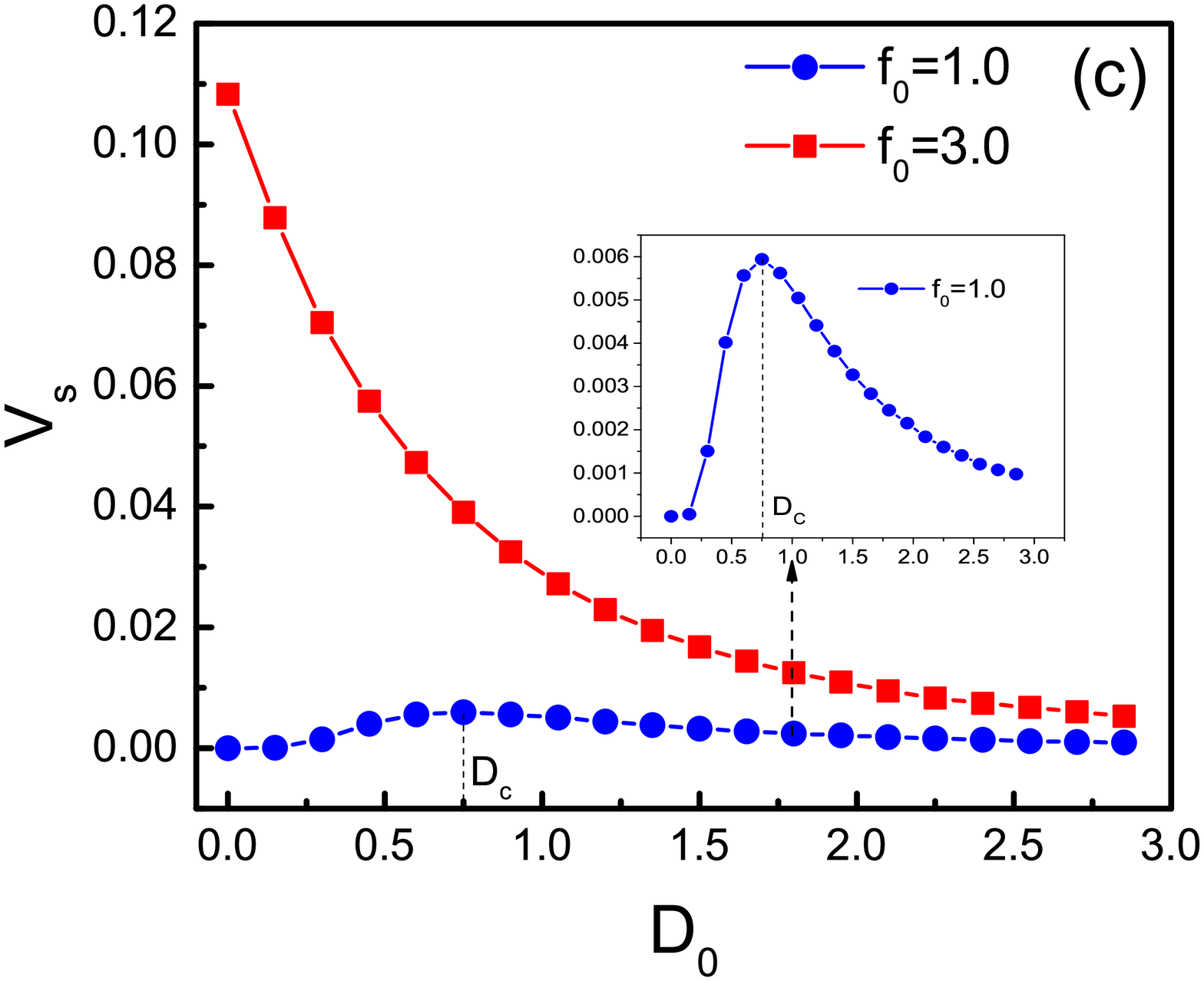}
  \includegraphics[width=0.45\columnwidth]{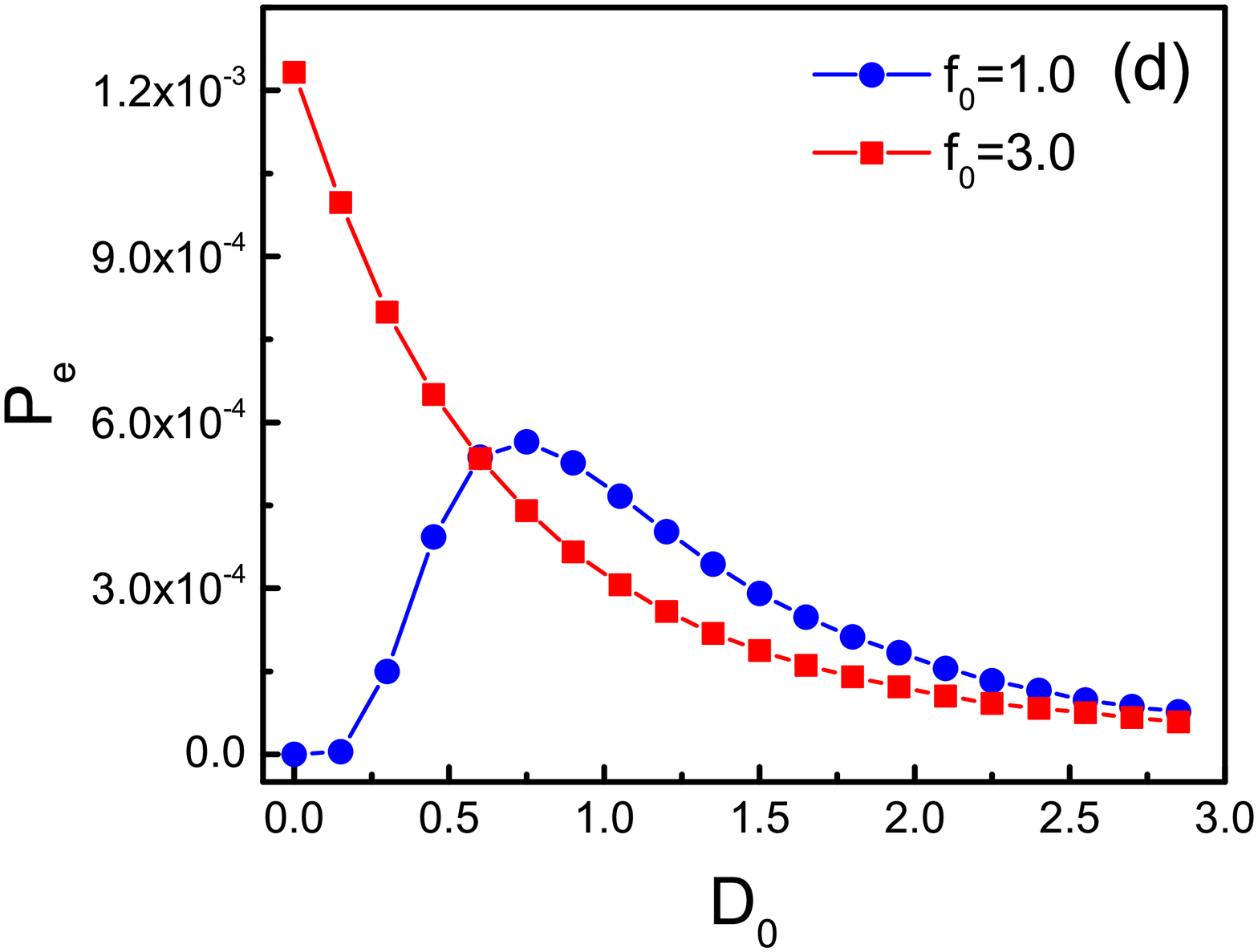}
  \caption{(a) (Color online)Average velocity $V_s$ as a function of $\Delta$ at $L_y=2.0$ and $D_0=0.0$. (b) Average velocity $V_s$ as a function of $L_y$ at $D_0=0.0$ and $\Delta=-\frac{1}{3}$. (c)Average velocity $V_s$ as a function of $D_0$ at $L_y=2.0$ and $\Delta=-\frac{1}{3}$. (d) P$\acute{e}$clet number $P_e$ as a function of $D_0$ at $L_y=2.0$ and $\Delta=-\frac{1}{3}$. The other parameters are $D_\theta=0.01$ and $\Gamma=3.2$.
    \label{fig:Parallel_pressure_current}}
\end{figure}

\indent Figure 7(a) shows the average velocity $V_s$ as a function of the asymmetric parameter $\Delta$ at $L_y=2.0$ and $D_0=0.0$.  It is found that when $f_0=3.0$ ($18.0$) the average velocity $V_s$ is positive (negative) for $\Delta<0$, zero at $\Delta=0$, and negative (positive)for $\Delta>0$. Obviously, the ratchet effect vanishes for the symmetric potential and no spontaneous symmetry breaking occurs.  Therefore, we can have the current reversal by changing the
sign of $\Delta$, the asymmetry of the potential.

\indent The dependence of the average velocity $V_s$ on the parameter $L_y$ is shown in Fig. 7(b) at $D_0=0.0$ and $\Delta=-\frac{1}{3}$. On increasing the parameter $L_y$ from zero, the average velocity $V_s$ increases quickly, and finally it reaches the saturated value when $L_y$ is greater than the the persistence length $l_p$. This can be explained as follows. In the present case, the persistence length of the active particle is $l_p=f_0/\Gamma D_\theta=93.75$.  Due to hard wall boundary conditions  in the $y$-direction, if the particle meet the wall, the angle $\theta $ changes randomly. When $L_y<l_p$, the smaller the value $L_y$ the faster the angle $\theta$ changes, the fast change of $\theta$ (equivalent to large $D_\theta$) reduces the ratchet transport. Therefore, the average velocity $V_s$ increases with $L_y$. When $L_y>l_p$, the role of $y$-direction disappears, so average velocity $V_s$ reaches the saturated value.

\indent Figure 7(c) describes the average velocity $V_s$ as a function of the translational diffusion $D_0$ at $L_y=2.0$ and $\Delta=-\frac{1}{3}$.  The translational diffusion can activate Brownian motion of particles or reduce the self-propelled driving. When $f_0=3.0$, the particle can easily pass across the barrier of the potential, the translational diffusion reduces the self-propelled driving, which blocks the ratchet transport. Therefore, the average velocity decreases with increase of $D_0$.  However, when $f_0=1.0$, there exists an optimal value of $D_0=D_c$ at which the average velocity takes its maximal value. This is because the particle can not pass across the barrier for small values of $D_0$ when $f_0=1.0$, the translational diffusion activates Brownian motion, which facilitates the ratchet transport. Therefore, the average velocity increases with $D_0$ when $D_0<D_c$.

\indent In order to compare the average velocity with diffusion, we use the P$\acute{e}$clet numbers $P_e =V_s L_x/D_{eff}$ to measure the ratchet transport,
where $D_{eff}=D_0+\frac{f^2_0}{\Gamma^2 D_\theta}$.  The P$\acute{e}$clet number $P_e$ as a function of the translational diffusion $D_0$ at $L_y=2.0$ and $\Delta=-\frac{1}{3}$ is shown in Fig. 7(d). Similarly to Fig. 7(c), the P$\acute{e}$clet number $P_e$ decreases with increase of $D_0$ for $f_0=3.0$ and there exists an optimal value of $D_0$ at which $P_e$ takes its maximal value for $f_0=1.0$.  For both cases, the P$\acute{e}$clet number $P_e$ is very small. Because the P$\acute{e}$clet number is the product of the Reynolds number and the Schmidt number, the Schmidt number must be very small in the present system with the high Reynolds number.

\section {Concluding remarks}
\indent We have numerically investigated the ratchet transport of underdamped noninteracting particles in the asymmetric potential. When the inertial term is neglected (in the overdmaped limit), particles always on average move to the easy direction (the slanted side) of the substrate asymmetry.  However, when the inertial term is included, the transport behaviors become complex. For very small values of $\Gamma$, the average velocity is always positive.  For very
large values of $\Gamma$, the system reduces to its overdamped version, the transport behaviors are completely identical for both underdamped and overdamped cases. The most important feature for the underdamped case is the occurrence of current reversals for intermediate values of $\Gamma$.  For intermediate values of $\Gamma$, on increasing the self-propulsion force $f_0$, the average velocity firstly increases to
its maximal value, then decreases, and finally reverses its direction. In addition, on increasing the friction coefficient $\Gamma$ from zero, the average velocity can change its direction serval times for the appropriate values of the self-propulsion force $f_0$.  In addition, for the damping case the average velocity decreases monotonously with increasing $D_\theta$ and for the very weak damping case there exists an optimal value of $D_\theta$ at which the average velocity is maximal. Current reversals can provide a tool for controlling and manipulating the motion of self-propelled particles in complex and crowded environments. We expect that our results open the way for novel rectification (or separation) devices of active matter based on underdamped dynamics.

\indent Though the present model is very simple, the results we have presented could open up a new aspect of active matter ratchet systems. This active ratchet could be realized in some real active matter systems at the high Reynolds number regime(also with low Schmidt number) such as sphere Janus particles in a dusty plasm, colloidal particles in air, granular matter in dilute systems, and so on. The present analysis can be easily extended to more complex situations in the future.  In particular, it is interesting to add collective effects to study competing effects of inertia versus interaction in terms of reversals. Another interesting extension is particle separation by using interacting inertial ratchets. Finally, it is also interesting to study  whether other non-dissipative dynamics, such as Magnus forces, could come into play.

\section*{Acknowledgements}
\indent This work was supported in part by the National Natural Science Foundation of China (Grants No. 11575064 and No. 11175067), the GDUPS (2016), and the Natural Science Foundation of Guangdong Province (Grants No. 2014A030313426 and No. 2016A030313433).

\end{document}